\documentclass[aps,prl,twocolumn, titlepage,showpacs]{revtex4}

\usepackage{graphicx}

\usepackage{dcolumn}

\usepackage{bm}

\begin{document}

\title{Viscosity calculated in simulations of strongly-coupled dusty plasmas with gas friction}

\author{Yan Feng}
\email{yan-feng@uiowa.edu}
\author{J. Goree}
\author{Bin Liu}
\affiliation{Department of Physics and Astronomy, The University
of Iowa, Iowa City, Iowa 52242, USA}

\date{\today}

\begin{abstract}

A two-dimensional strongly-coupled dusty plasma is modeled using
Langevin and frictionless molecular dynamical simulations. The
static viscosity $\eta$ and the wave-number-dependent viscosity
$\eta(k)$ are calculated from the microscopic shear in the random
motion of particles. A recently developed method of calculating
the wave-number-dependent viscosity $\eta(k)$ is validated by
comparing the results of $\eta(k)$ from the two simulations. It is
also verified that the Green-Kubo relation can still yield an
accurate measure of the static viscosity $\eta$ in the presence of
a modest level of friction as in dusty plasma experiments.

\end{abstract}

\narrowtext

\maketitle

\section {I.~INTRODUCTION}

Strongly-coupled plasma is a collection of free charged particles
where the Coulomb interaction with nearest neighbors is so strong
that particles do not easily move past one another. A widely used
criterion to determine whether a plasma is strongly coupled is
$\Gamma>1$~\cite{Ichimaru:82, Fortov:06}, where $\Gamma$ is
defined as the ratio of the potential energy between neighboring
particles and the kinetic energy. When $\Gamma>1$, particles move
slowly and are trapped by a cage consisting of a few nearby
particles. If they escape the cages gradually, particles in a
strongly-coupled plasma can flow, much like a
liquid~\cite{Vieillefosse:75}. However, if $\Gamma\gg10^2$, nearby
particles that form a cage move so little that a particle inside
the cage can seldom escape the cage; this condition is like
molecules in a solid~\cite{Donko:02, Io:09}. If a shearing stress
is applied, cages in a solid are elastically deformed but can
restore to their previous state, whereas cages in a liquid are
disrupted and a viscous flow can develop.

One type of strongly-coupled plasma is dusty plasma formed in the
laboratory. A dusty plasma consists of four constituents:
micron-size particles of solid matter (dust particles), electrons,
ions, and neutral gas atoms~\cite{Morfill:09, Piel:10, Shukla:02}.
The dust particles are strongly coupled amongst themselves due to
a large interparticle potential energy provided by a large
particle charge~\cite{Merlino:04, Feng:08}. Several schemes have
been used to confine charged dust particles using natural electric
field inside a plasma. One of these schemes makes use of a
radio-frequency plasma~\cite{Chu:94, Gavrikov:05}, with a
horizontal electrode that provides a sheath electric field that
can confine and levitate dust particles in a cloud with only a few
horizontal layers. If experimenters introduce only a limited
number of dust particles, they can settle into just a single
layer~\cite{Feng:08}. In these single-layer clouds, dust particles
have negligible vertical motion, so that the cloud of dust
particles is often described as a two-dimensional (2D)
system~\cite{Melzer:96, Nunomura:00, Wang:01, Nosenko:04,
Feng:08}. In this 2D cloud, the interaction between dust particles
is a repulsive Yukawa potential~\cite{Konopka:00}. Due to the
large length scale and the slow time scales~\cite{Morfill:09},
dusty plasmas allow video microscopy to track individual particle
motion~\cite{Feng:07}. In dusty plasma experiments, elasticity in
solids~\cite{Nunomura:00} and viscosity in
liquids~\cite{Nosenko:04} has been observed and studied. However,
strongly-coupled plasmas cannot always be classified as purely
elastic or purely viscous.

Dust particles experience several forces in the experiments. The
electric force provides strong coupling amongst the dust particles
as well as the levitation and confinement. Gas friction, due to
dust particles moving relative to the rarefied gas, is the primary
energy loss mechanism. The gas is usually so rarefied that it
represents only a small portion of the mass of the dusty plasma.
Gas represents $<10\%$ of the mass of dust in a 3D dusty plasma
experiment at $400~{\rm mTorr}$~\cite{Flanagan:09}, while 2D
experiments have even less gas, with a pressure $< 20~{\rm
mTorr}$~\cite{Feng:10, Feng:10_3, Feng:08}. There is an ion drag
force due to a steady flow of ions, arising from the same dc
electric fields that provide levitation and confinement of dust
particles. This ion drag force is parallel to the ion flow.
Finally, in some experiments, laser radiation pressure forces are
used to accelerate dust particles, for example to create
macroscopic flows~\cite{Juan:01, Nosenko:04, Gavrikov:05,
Feng:10_3} or simply raise the kinetic temperature of the dust
particles without causing a macroscopic flow~\cite{Wolter:05,
Nosenko:06, Feng:08, Feng:10}. This kind of laser heating method
is one of several ways that experimenters can control $\Gamma$ so
that the cloud of dust particles behaves like a liquid or
solid~\cite{Nosenko:06, Samsonov:04, Sheridan:08, Thomas:96,
Melzer:96}.

We assume that the Coulomb interaction amongst charged dust
particles is the dominant mechanism for viscosity in laboratory
dusty plasma experiments. Viscous transport of momentum occurs
when the dust particles moving relative to one another in a
shearing motion collide, causing some of their momentum to be
transferred across the flow. We expect that collisions involving
gas atoms will contribute less to the viscosity. Although the
force of gas friction is effective in diminishing momentum of dust
particles in the direction of their motion, there are two reasons
it has little effect in transferring momentum across a flow of
dust particles. First, the gas is rarefied so that it can carry
much less momentum than a viscous solvent in a colloidal
suspension~\cite{Einstein:06, Einstein:11}, for example. Second,
in a 2D experiment like~\cite{Feng:10}, a gas atom that is struck
by a dust particle is usually knocked into a direction out of the
dust layer, so that there is little opportunity for a dust
particle to push another dust particle indirectly through
collisions with a gas atom.

Here, we will refer to the viscosity as the static viscosity
$\eta$ to distinguish it from viscoelasticity. In the literature
of dusty plasmas, the static viscosity $\eta$ has been measured
experimentally~\cite{Nosenko:04} and quantified in
simulations~\cite{Saigo:02, Liu:05, Donko:06, Donko:08}. There are
two ways to quantify the static viscosity. If there is a
macroscopic velocity shear, the static viscosity can be calculated
from the velocity flow profile~\cite{Nosenko:04, Gavrikov:05,
Donko:06, Donko:08}. On the other hand, if there is no macroscopic
velocity shear, the microscopic shear associated with the random
motion of particles can be used to calculate the static viscosity
using the Green-Kubo relation~\cite{Liu:05, Donko:08, Donko:09}.

Viscoelasticity is a property of materials that exhibit both
liquid-like viscous and solid-like elastic
characteristics~\cite{Lakes:09}. Most materials in reality are
viscoelastic, such as wood, synthetic polymers, and human
tissue~\cite{Lakes:09}. Viscous effects correspond to energy
dissipation, while elastic effects corresponds to energy storage.
In general, liquids exhibit mostly viscous effects at large
spatial and temporal scales, but they exhibit some elastic effects
at small spatial and temporal scales~\cite{Ratynskaia:09}.

To characterize viscoelasticity quantitatively, it is common to
use either the frequency-dependent viscosity $\eta(\omega)$ or the
wave-number-dependent viscosity $\eta(k)$. The latter
characterizes materials at different length scales, and was
introduced by theorists performing simulations~\cite{Balucani:87,
Palmer:94, Balucani:00, Hu:07}. The static viscosity $\eta$ is the
hydrodynamic limit of the wave-number-dependent viscosity
$\eta(k)$ when $k \rightarrow 0$. In considering this limit, the
relevant characteristic length scale for $k$ is the interparticle
distance, which is often measured as the lattice constant $b$ of a
perfect crystal.

Viscoelasticity of strongly-coupled plasmas has been studied
theoretically~\cite{Kaw:98, Murillo:00, Donko:10} and
experimentally~\cite{Ratynskaia:06, Chan:07, Feng:10}. The few
experiments that have been reported for viscoelasticity of dusty
plasma include a descriptive presentation~\cite{Ratynskaia:06} and
a characterization using a correlation function of the microscopic
motion of dust particles~\cite{Chan:07}.

In our recent 2D experiment~\cite{Feng:10}, a single horizontal
layer of electrically charged dust particles was levitated in a
glow-discharge plasma. The kinetic temperature of the dust cloud
was raised by laser heating~\cite{Feng:10, Nosenko:06}. Viewing
from above, we recorded movies of particle motion, then calculated
particle positions and tracked them to calculate their velocities.
Based on the trajectories of particles, the wave-number-dependent
viscosity $\eta(k)$ of the 2D dusty plasma was quantified using an
expression we derived that accounts for gas friction.

In simulations, the viscoelasticity of both 2D~\cite{Feng:10} and
3D strongly-coupled plasmas~\cite{Donko:10} have been studied
recently. In this paper, we carry out further simulations for two
purposes: to validate the $\eta(k)$ calculation method taking into
account gas friction, as presented in~\cite{Feng:10}, and to
assess the accuracy of the Green-Kubo relation for dusty plasmas
with a modest level of gas friction.

Simulations of strongly-coupled plasmas usually use the molecular
dynamical (MD) method~\cite{Feng:10, Donko:10}. Each particle is
tracked individually, unlike the case of particle-in-cell (PIC)
simulations, where aggregations of particles are simulated by a
hypothetical super-particle~\cite{Piel:10}. Tracking individual
particles is suitable because otherwise the dominant effects of
strong particle-particle Coulomb interactions would be lost.
Another difference is that in MD simulations, as compared to PIC
simulations, Poisson's equation is not solved. The only equation
that is solved is the equation of motion for each particle, which
is integrated to track particle trajectories. The result of the MD
simulation is a record of all particle positions and velocities,
which is the same kind of data that are produced in dusty plasma
experiments. The interparticle interaction that is assumed in MD
simulations of strongly-coupled dusty plasmas is a repulsive
Yukawa potential~\cite{Konopka:00},
\begin{equation}\label{Yukawa}
\phi_{i,j}=Q^2(4\pi\epsilon_0r_{i,j})^{-1}{\rm
exp}(-r_{i,j}/\lambda_D),
\end{equation}
where $Q$ is the charge on dust particles, $\lambda_D$ is the
Debye length, and $r_{i,j}$ is the distance between the $i$th and
$j$th particles.

We list here additional parameters for the dusty plasma cloud.
Because the dust cloud is 2D, we use an areal number density $n$
and an areal mass density $\rho = mn$ for the cloud, where $m$ is
the dust particle mass. We note that while the units for mass
density and viscosity are different in 2D and 3D, the units are
the same for the kinematic viscosity~\cite{Nosenko:04},
$\eta/\rho$. Distances between dust particles are characterized by
both the lattice constant $b$ for a crystal or the 2D Wigner-Seitz
radius $a = (n \pi)^{-1/2}$~\cite{Kalman:04}. Time scales for
collective motion are characterized by the nominal 2D dusty plasma
frequency~\cite{Kalman:04} $\omega_{pd} = (Q^2/2\pi\epsilon_0 m
a^3)^{1/2}$. Gas friction is characterized by the damping rate
$\nu_f$, which is the ratio of the gas friction force and the dust
particle's momentum.

We will discuss how to calculate $\eta$ and $\eta(k)$ in Sec.~II.
In Sec.~III, we will discuss our two MD simulation methods,
Langevin and frictionless. In Sec.~IV, we will report new
simulation data for $\eta(k)$ of 2D strongly-coupled dusty
plasmas. We will validate our analysis method~\cite{Feng:10} for
calculating $\eta(k)$ in 2D strongly-coupled plasma with gas
friction. We will also test the accuracy of the Green-Kubo
relation with a modest level of gas friction as in our experiment.

\section {II.~METHODS FOR CHARACTERIZING VISCOSITY}

\subsection {A. Static viscosity $\eta$}
The Green-Kubo relation is widely used for calculating the static
viscosity $\eta$, based on the random motion of particles. This
method is used when there is no macroscopic velocity shear. The
Green-Kubo approach assumes linear microscopic fluctuations and
equilibrium fields in the system~\cite{Liu:05}. The assumptions of
this approach are similar to those for the fluctuation-dissipation
theorem~\cite{Montgomery:64, Pathria:72}. Previously, the
Green-Kubo relation was generally used with data from frictionless
simulations~\cite{Liu:05, Donko:10, Donko:09, Donko:08, Saigo:02}.
To calculate the static viscosity, first we calculate the stress
autocorrelation function (SACF)
\begin{equation}\label{SACF}
{C_{\eta}(t)= \langle P_{xy}(t)P_{xy}(0) \rangle,}
\end{equation}
where $P_{xy}(t)$ is the shearing stress
\begin{equation}\label{SS}
{P_{xy}(t)=
\sum_{i=1}^N\left[mv_{ix}v_{iy}-\frac{1}{2}\sum_{j\not=i}^N\frac{x_{ij}y_{ij}}{r_{ij}}\frac{\partial
\phi(r_{ij})}{\partial r_{ij}}\right],}
\end{equation}
where $i$ and $j$ are indices for different particles, $N$ is the
total number of particles of mass $m$, $\mathbf{r}_{i} =
(x_i,y_i)$ is the position of particle $i$, $x_{ij}=x_i-x_j$,
$y_{ij}=y_i-y_j$, $r_{ij}=|\mathbf{r}_i-\mathbf{r}_j|$, and
$\phi(r_{ij})$ is the interparticle potential. Second, we
calculate the static viscosity $\eta$ from the Green-Kubo
relation~\cite{Liu:05},
\begin{equation}\label{eta}
{\eta=\frac{1}{VkT}\int^\infty_0 C_{\eta}(t)dt.}
\end{equation}
Here, $V$ is the simulation volume, which is replaced by the area
of the simulation box for 2D simulations like those reported here.

The Green-Kubo relation, Eq.~(\ref{eta}), is intended for use in
equilibrium systems, but in this paper we will assess whether it
can also be used in systems with a modest level of gas friction as
in our experiment~\cite{Feng:10}. The dust particles in an
experiment experience gas friction, in addition to collisions
amongst themselves, whereas only the latter are modeled in the
Green-Kubo relation. We will carry out simulations, with and
without friction, and verify that Eq.~(\ref{eta}) yields the same
result in both cases.

\subsection {B. Wave-number-dependent viscosity $\eta(k)$}
The wave-number-dependent viscosity $\eta(k)$ characterizes
viscous effects at different length scales. A method of
calculating $\eta(k)$ from the trajectories of random motion of
molecules in liquids has been developed~\cite{Balucani:00, Hu:07}.
In calculating $\eta(k)$ using this method, one starts with
particle trajectories, such as $x_i(t)$ and the perpendicular
velocity $v_{iy}(t)$ for the $i$th particle. These are used to
calculate the transverse current, $j_y(k,t)=\sum\nolimits^N_{i=1}
v_{iy}(t)\,{\rm exp}[ikx_i(t)]$. The normalized transverse current
autocorrelation function~\cite{Balucani:00, Hu:07} (TCAF) is then
calculated as
\begin{equation}\label{TCAF}
{C_T(k,t) = \langle j^*_y(k,0)\,j_y(k,t)\rangle / \langle
j^*_y(k,0)\,j_y(k,0)\rangle ,}
\end{equation}
where the wave vector $k$ is parallel to the $x$ axis. (Here, $k$
serves only as a Fourier transform variable, and is not intended
to characterize any waves.) The wave-number-dependent viscosity of
frictionless systems is calculated~\cite{Balucani:00, Hu:07} using
\begin{equation}\label{etakold}
{\eta(k)/\rho=1/(\Phi k^2),}
\end{equation}
where $\Phi$ is a time integral representing the area under the
TCAF after normalizing the TCAF to have a value of unity at $t=0$.
Generally, $\eta(k)$ diminishes gradually as $k$ increases,
meaning that viscous effects gradually diminish at shorter length
scales.

In~\cite{Feng:10}, we generalized this expression as
\begin{equation}\label{etak}
{\eta(k)/\rho=\left[(1/\Phi)-\nu_f\right]/k^2}
\end{equation}
to account for the friction of gas drag $\nu_f$ acting on dust
particles. As in Eq.(\ref{etakold}), the integral $\Phi$ is a
function of $k$. Our derivation of Eq.~(\ref{etak}) was provided
in the supplementary material of~\cite{Feng:10}. In this paper, we
will carry out simulation tests to validate the use of
Eq.~(\ref{etak}) for a wide range of $k$. This validation test
will be performed for the modest level of gas friction $\nu_f$ in
our experiment~\cite{Feng:10}.

The TCAF measures the memory of transverse current, which reflects
the decay of microscopic velocity shear. The shear decay can be
caused by several mechanism in 2D dusty plasma clouds, such as
Coulomb collisions amongst dust particles and the friction due to
gas drag. We will study how gas friction affects the TCAF later.

\section {III.~SIMULATION METHODS}

In order to test the effects of gas friction, we will compare the
results of two simulations: a Langevin MD simulation with
friction, and a frictionless equilibrium MD simulation.

Our two simulation methods are the same in many respects. Both use
a binary interparticle interaction with a Yukawa pair potential.
In both simulations, particles are only allowed to move in a
single 2D plane. Conditions remained steady during each simulation
run. For both simulations, the parameters we used were $N = 4096$
particles in a rectangular box with periodic boundary conditions.
The box had sides $64.1b\times55.5b$. The integration time step
was $0.019~\omega_{pd}^{-1}$, and simulation data were recorded
for a time duration of $68~000~\omega_{pd}^{-1}$ after a steady
state was reached. Both of our simulations were performed at
$\Gamma=68$ and $\kappa=0.5$, which are the same values as in our
experiment~\cite{Feng:10}.

Our Langevin MD simulation takes into account the dissipation due
to gas friction. The equation of motion that is integrated in the
Langevin simulation is~\cite{Feng:08_2, Feng:10_2, Donko:10,
Hou:09, Ott:09, Vaulina:09, Ratynskaia:09}
\begin{equation}\label{LDmotion}
m\ddot{\mathbf{r}}_{i}=-\nabla \sum \phi_{ij}-\nu_f
m\dot{\mathbf{r}}_{i}+\zeta_{i}(t),
\end{equation}
where $\nu_f m\dot{\mathbf{r}}_{i}$ is a frictional drag and
$\zeta_{i}(t)$ is a random force. There is no thermostat to adjust
the temperature; instead the temperature is established by
choosing the magnitude of $\zeta_{i}(t)$. Here, we chose the
experimental value $\nu_f=0.08~\omega_{pd}$~\cite{Feng:10}. Note
that this gas friction level is modest, i.e., the dust particle
motion is underdamped, since $\nu_f\ll \omega_{pd}$.

Our frictionless equilibrium MD simulation~\cite{Feng:10_2,
Donko:10, Liu:07} has no gas friction in the equation of motion
\begin{equation}\label{MDmotion}
m\ddot{\mathbf{r}}_{i}=-\nabla \sum \phi_{ij}.
\end{equation}
A Nos\'e-Hoover thermostat is applied to maintain a desired
temperature \cite{Liu:07, Feng:10_2}.

Trajectories $\mathbf{r}_i(t)$ are found by integrating
Eq.~(\ref{LDmotion}) or (\ref{MDmotion}) for all particles. An
example is shown in Fig.~1 from the frictionless MD simulation.

\section {IV.~RESULTS}

\subsection {A. Hydrodynamic and viscoelastic regimes}

Comparing the results from the two simulations, Fig.~2, we can see
how friction speeds the loss of memory of the system's microscopic
shearing motion. The memory of the shearing motion is indicated by
the decay of the TCAF.

As expected~\cite{Feng:10}, in the typical hydrodynamic limit of
long length scales, as shown in Fig.~2(a), the TCAF is just a
monotonic decay from unity to zero without any
oscillations~\cite{Feng:10}. We find that at the same hydrodynamic
length scale, the TCAF decays much faster with friction than
without, indicating that in experimental dusty plasmas gas
friction plays an important role in shear decay in large length
scales.

When the wave number $k$ is slightly larger, in the intermediate
regime between the hydrodynamic and viscoelastic regimes,
Fig.~2(b), the difference in TCAF between frictional and
frictionless is smaller. The integral of the frictional TCAF is
about a half of that for the frictionless TCAF, as seen in the
inset of Fig.~2(b). This integral corresponds to $\Phi$, as in
Eq.~(\ref{etakold}) or Eq.~(\ref{etak}).

When the wave number $k$ is even larger, in the viscoelastic
regime, the TCAF oscillates around zero after its decay due to the
elastic effects, Fig.~2(c). In this viscoelastic regime, there is
little difference between the TCAF from the two simulations,
indicating that at smaller length scales, gas friction does not
contribute much to shear decay. The friction plays a larger role
in TCAF at larger length scales than smaller length scales.

The calculation of $\eta(k)$ using Eq.~(\ref{etakold}) or
(\ref{etak}) requires choosing an upper limit in the time integral
of TCAF $C_T(k,t)$. An infinite time is of course impractical for
both experiments and simulations, so for a finite value we chose
$t_I$, the time of the first upward zero crossing of the
TCAF~\cite{Feng:10}, as shown in Fig.~2(c). This choice is
suitable for two reasons: first, it is sufficiently long to retain
both viscous and elastic effects; second, we found that
contributions to the integral after $t_I$ are negligible, for a
TCAF that is not noisy. The calculation result for $\eta(k)$ is
not very sensitive to the chosen upper limit. Extending the limit
to a higher value would only cause a limited effect on the value
of the integral.

\subsection {B. Validating the generalized $\eta(k)$ expression}

Results for the wave-number-dependent viscosity $\eta(k)$ are
presented in Fig.~3(b) and (c) for both simulations.

We find an agreement in the values of $\eta(k)$ for the
frictionless and Langevin simulations. This agreement can be seen
by comparing the circles in Fig.~3(b) for the frictionless
simulation with Eq.~(\ref{etakold}), and the triangles in
Fig.~3(c) for the Langevin simulation with Eq.~(\ref{etak}). There
is not only a qualitative agreement in the downward trend as the
wave number $k$ increases, but also a quantitative agreement. This
quantitative agreement is most easily seen by fitting the
calculated $\eta(k)$ to the Pad\'e approximant of~\cite{Feng:10,
Balucani:00} and comparing the fit parameters, as indicated in
Fig.~3 for the smooth curves.

This agreement leads us to our first chief result: a validation of
Eq.~(\ref{etak}) for computing $\eta(k)$ in the presence of gas
friction. Since the two simulations were performed for the same
values of $\Gamma$ and $\kappa$, an agreement indicates that
Eq.~(\ref{etak}) is valid. If there had been a discrepancy between
the circles in Fig.~3(b) and the triangles in Fig.~3(c), we would
question whether Eq.~(\ref{etak}) is valid. We gain confidence in
the validity of Eq.~(\ref{etak}) by the lack of any significant
discrepancy in the two results.

The importance of correcting for friction, in Eq.~(\ref{etak}), is
demonstrated in Fig.~3(c). If we use Eq.~(\ref{etakold}) instead,
the presence of friction leads to an exaggerated value for
$\eta(k)$, as seen by comparing the two sets of data in Fig.~3(c).
This exaggeration is most extreme at small wave numbers (where the
effect of friction is greatest, as we found in Sec.~IV~A for the
TCAF).

\subsection {C. Testing the Green-Kubo relation for static viscosity in the presence of friction}

To determine whether the Green-Kubo relation, Eq.~(\ref{eta}),
still provides an accurate calculation of static viscosity $\eta$
of a 2D Yukawa liquid, in the presence of a modest level of gas
friction, we performed a test of Eq.~(\ref{eta}) comparing $\eta$
computed from our frictional Langevin simulation and our
frictionless simulation. These results for the normalized
kinematic static viscosity are
$\eta/\rho=(0.26\pm0.02)~a^2\omega_{pd}$ for the frictionless
simulation, and $\eta/\rho=(0.27\pm0.02)~a^2\omega_{pd}$ for the
Langevin simulation with friction. These values are also shown in
Fig.~3(b) and (c) as star symbols. Noting that these results are
in agreement within the uncertainty, we conclude that the
Green-Kubo relation remains accurate, at least with a modest level
of gas friction, for a 2D Yukawa liquid at the value $\Gamma=68$
and $\kappa=0.5$.

A further confirmation of the accuracy of the Green-Kubo relation
when used with modest levels of friction can be found by examining
our $\eta(k)$ in Fig.~3(c). We note an agreement of  $\eta(k)$ as
$k \rightarrow 0$ with $\eta$ from the Green-Kubo relation. This
agreement is significant because $\eta(k)$ is computed from the
TCAF, which is unrelated to the Green-Kubo relation used to
compute $\eta$.

We can provide two intuitive suggestions to explain the accuracy
of the Green-Kubo relation in the presence of a modest level of
gas friction. First, we note that the gas friction that we have
considered is so small that $\nu_f/\omega_{pd}<0.1$. This
inequality demonstrates that frictional effects will in general be
much smaller than effects arising from particle charge as measured
by $\omega_{pd}$. Second, the TCAF in Fig.~2 showed us that gas
friction has the least effect on motion at small length scales,
and dynamical information at these small length scales are also
reflected in the Green-Kubo relation because it is based only on
fluctuations of individual particle motion.

We cannot rule out the possibility that friction will affect the
static viscosity computed using the Green-Kubo relation in other
parameter regimes. In fact, for a 3D Yukawa Langevin simulation at
a much lower $\Gamma=2$, Ramazanov and Dzhumagulova found that
$\eta$ computed using the Green-Kubo relation diminishes as the
friction was raised to a very high level~\cite{Ramazanov:08}.

\section {IV.~Summary}

Motivated by experiments with 2D clouds of charged dust particles
suspended in a plasma, we carried out two types of simulations,
with and without gas friction. We validated the newly-introduced
Eq.~(\ref{etak}) for calculating $\eta(k)$ as a measure of
viscoelasticity, in the presence of gas friction. We also verified
that the Green-Kubo relation can accurately measure the static
viscosity $\eta$ of the 2D collection of charged dust particles
even when they experience gas friction. The level of gas friction
we considered was at a low level $\nu_f/\omega_{pd}<0.1$, and the
coupling was moderate with $\Gamma=68$ and $\kappa=0.5$, both as
in our recent experiment~\cite{Feng:10}.

This work was supported by NSF and NASA.

\begin{figure}[p]
\caption{\label{MDTrajectories} (color online). Trajectories of
particles from the frictionless MD simulation. Similar
trajectories for the experiment and a Langevin MD simulation were
reported in~\cite{Feng:10}. All simulation results here are for
$\Gamma=68$, $\kappa=0.5$.}
\end{figure}

\begin{figure}[p]
\caption{\label{TCAFs} (color online). Transverse current
autocorrelation functions (TCAF) from simulations in different
regimes: (a) hydrodynamic, (b) intermediate between hydrodynamic
and viscoelastic, and (c) viscoelastic. Friction plays a larger
role at smaller $k$, i.e., longer length scales. After a long
time, the TCAF always decays to zero. (For the small $k$ case
without friction (a), the TCAF approaches zero after a great time,
longer than shown here.) The inset in (b) is the cumulative time
integral of TCAF.}
\end{figure}

\begin{figure}[p]
\caption{\label{eta-k} (color online). Wave-number-dependent
viscosity $\eta(k)$ from (a) the experiment, (b) the Langevin MD
simulation, and (c) the frictionless MD simulation. The agreement
of the smooth curves in (b) and (c) validates Eq.~(\ref{etak}) for
calculating $\eta(k)$ in the presence of gas friction. Using
Eq.~(\ref{etakold}) in the presence of gas friction (circle data
points in (c)) would lead to an exaggerated value for $\eta(k)$,
especially at longer length scales. Also shown in (b) and (c) is
the calculated static viscosity $\eta$ based on the Green-Kubo
relation as indicated with star symbols. Panel (a) is reprinted
from \cite{Feng:10}.}
\end{figure}

\end{document}